\documentclass[twocolumn,showpacs,preprintnumbers,amsmath,amssymb]{revtex4}

\usepackage{graphicx}
\usepackage{dcolumn}
\usepackage{bm}
\newtheorem{theorem}{Theorem}[section]
\newtheorem{proposition}[theorem]{Proposition}

\newtheorem{corollary}[theorem]{Corollary}
\newtheorem{definition}[theorem]{Definition}
\newtheorem{remark}[theorem]{Remark}
\newtheorem{conjecture}[theorem]{Conjecture}
\newcommand{\eproof}{\rule{0,2cm}{0,2cm}}

\begin{document}

\preprint{APS/123-QED}

\title{Generalization of symmetric $\alpha$-stable L\'evy distributions for $q > 1$}

\author{Sabir Umarov}
\affiliation{Department of Mathematics, Tufts University, Medford, MA 02155, USA}%
\email{Sabir.Umarov@tufts.edu}

\author{Constantino Tsallis}%
\affiliation{Centro Brasileiro de Pesquisas Fisicas and National Institute of Science and Technology for Complex Systems\\
Rua Dr. Xavier Sigaud 150, 22290-180 Rio de Janeiro, RJ, Brazil\\
and\\
Santa Fe Institute, 1399 Hyde Park Road, Santa Fe, NM 87501, USA}
 \email{tsallis@cbpf.br}

\author{Murray Gell-Mann}
\affiliation{Santa Fe Institute, 1399 Hyde Park Road, Santa Fe, NM 87501, USA
}%
\email{mgm@santafe.edu}

\author{Stanly Steinberg}
\affiliation{Department of Mathematics and Statistics, University of New Mexico, Albuquerque, NM 87131, USA}%
\email{stanly@math.unm.edu}


\begin{abstract} The $\alpha$-stable distributions introduced by L\'evy  play an important role in probabilistic
theoretical studies and their various applications, e.g.,  in
statistical physics, life sciences, and economics. In the present
paper we study sequences of long-range dependent random variables
whose distributions have asymptotic power law decay, and which are
called $(q,\alpha)$-stable distributions. These sequences are
generalizations of i.i.d. $\alpha$-stable distributions, and have not been previously
studied. Long-range dependent $(q,\alpha)$-stable distributions
might arise in the description of anomalous processes in nonextensive
statistical mechanics, cell biology, finance. The parameter $q$
controls dependence. If $q=1$ then they are classical i.i.d. with
$\alpha$-stable L\'evy distributions. In the present paper we establish
basic properties of $(q,\alpha)$-stable distributions, and
generalize the result of Umarov, Tsallis and Steinberg (2008), where the
particular case $\alpha=2, \, q\in [1,3),$ was considered, to the
whole range of stability  and nonextensivity parameters $\alpha \in
(0,2]$ and $q \in [1,3),$ respectively. We also discuss possible
further extensions of the results that we obtain, and formulate some
conjectures. \end{abstract}

\maketitle

\section{Introduction}
\label{intro} The central limit theorem (CLT) and $\alpha$-stable
distributions have rich applications in various fields including the
Boltzmann-Gibbs (BG) statistical mechanics.
The nonextensive statistical mechanics
\cite{Tsallis1988,PratoTsallis1999,GellmannTsallis,BoonTsallis,encyclopedia,Tsallis2009}
characterized by the nonextensivity index  $q$ (which recovers the
BG theory in the case $q=1$) studies, in particular, strongly
correlated random states, mathematical models of which can be
represented by specific long-range dependent random variables.
The $q$-central limit theorem consistent with nonextensive
statistical mechanics was established in paper
\cite{UmarovTsallisSteinberg}. The main objective of
\cite{UmarovTsallisSteinberg} was to study
the scaling limits (attractors) of sums of $q$-independent random
variables
with a finite $(2q-1)$-variance. The
mapping
\begin{equation}
\label{paper1} F_q: \mathcal{G}_q [2] \rightarrow
\mathcal{G}_{z(q)}[2] \,,
\end{equation}
where $F_q$ is the $q$-Fourier transform  (see Section 2) $
z(s)={(1+s)}/{(3-s)},$ and $\mathcal{G}_q[2]$ is the set of
$q$-Gaussians up to a constant factor (see, e.g.
\cite{PratoTsallis1999,GellmannTsallis}), was essentially used in
the description of attractors. The number 2 in the notation will
soon become transparent in the context of the current paper.

In the present work we study a $q$-analog of the $\alpha$-stable
L\'evy distributions. In this sense, the present paper is a
conceptual continuation of \cite{UmarovTsallisSteinberg}.
The classic theory of $\alpha$-stable distributions
was originated by Paul L\'evy and developed by
L\'evy, Gnedenko, Feller and others; for details and history see,
for instance, [8-14] and references therein. Distributions with
asymptotic power-law decay ($\alpha$-stables, and particularly
$q$-Gaussians, in the first place) found a huge number of
applications in various practical studies (see, e.g., [14-23], just
to mention a few), confirming the frequent nature of these
distributions. As it will become clear later on, $(q,\alpha)$-stable
distributions  unify both of them. Indeed, $(1,\alpha)$-stable
distributions correspond to the $\alpha$-stable ones, and the
$(q,2)$-stable distributions correspond to the $q$-Gaussian ones.
All $(q,\alpha)$-stable distributions, except Gaussians
($(1,2)$-distributions), exhibit asymptotic power-laws. In practice
the researcher is often interested in identification of a correct
attractor of correlated states, which plays a major role in the
adequate modeling of physical phenomenon itself. This motivates the
study of sequences of $(q,\alpha)$-stable distributions and their
attractors, as focused in the present paper.

For simplicity we will consider only symmetric $(q,\alpha)$-stable
distributions in the one-dimensional case (see
\cite{UmarovTsallis2007} for the multivariate $q$-CLT). We denote
the class of random variables with $(q,\alpha)$-stable distributions
by $\mathcal{L}_q[\alpha].$  A random variable $X \in
\mathcal{L}_q[\alpha]$ has a symmetric density $f(x)$ with
asymptotics $f \sim C |x|^{-\frac{1+\alpha}{1+\alpha(q-1)}}, \, \,
|x|\rightarrow \infty$, where $1\le q <2, ~ 0<\alpha < 2,$ and $C$
is a positive constant. {Hereafter $g(x) \sim h(x), x \rightarrow
a,$ means that $\lim_{x \rightarrow a}\frac{g(x)}{h(x)} = 1$.}
Linear combinations and properly scaling limits of sequences of
$q$-independent random variables with $(q,\alpha)$-stable
distributions are again random variables with $(q,\alpha)$-stable
distributions, justifying that $\mathcal{L}_q[\alpha]$ form a class
of "stable" distributions. To this end, we note that
$\mathcal{L}_q[\alpha]$
shares the same asymptotic behavior with the set $\mathcal{L}_{sym}(\gamma)$ of symmetric
L\'evy distributions centered at $0,$  where
$$\gamma=\gamma(q,\alpha)=\frac{\alpha (2-q)}{1+\alpha(q-1)}.$$
However, there is an essential difference between
$(q,\alpha)$-stability of $q$-independent random variables and the
classic stability of $\alpha$-stable distributions. Namely,
$q$-independence exhibits a special long-range correlation between
random variables (see the exact definition in Section 3). In
practice this notion reflects  physical states (arising e.g. in
nonextensive statistical mechanics), which are strongly correlated.
The term "global correlation" instead of "strong correlation" is
also used widely in physics literature; see, e.g.
\cite{GellmannTsallis,BoonTsallis}. Examples of such systems include
earthquakes \cite{Caruso}, cold atoms in optical dissipative
lattices \cite{Renzoni}, and dusty plasma \cite{Goree}. A
decomposition of nonextensive processes with strong correlation into
independent states can not adequately reflect their evolution.
Likewise, $(q,\alpha)$-stable distributions can not be captured
by the existing theory of $\alpha$-stable distributions, which is
heavily based on the concept of independence (or weak dependence). This distinction ends
up with essential implication: attractors of $(q,\alpha)$-stable
distributions are different from the attractors of $\alpha$-stable
distributions, unless $q=1.$ If $q=1$ then correlation disappears,
that is $q$-independence becomes usual probabilistic independence,
and $\gamma(1,\alpha)=\alpha,$ implying
$\mathcal{L}_1[\alpha]=\mathcal{L}_{sym}[\alpha].$
\par
Following the method established in \cite{UmarovTsallisSteinberg},
we will apply $F_q$-transform for the study of sequences of
$q$-independent $(q,\alpha)$-stable distributions.
Parameter $q$ controls correlation.
We will classify $(q,\alpha)$-stable distributions
depending on parameters $1 \le q < 2$ (or equivalently $1 \le Q<3$,
$Q=2q-1$) and $0<\alpha\leq 2$. We establish the mapping
\begin{equation}
\label{paper2part1} F_q: \mathcal{G}_{q^{L}} [2] \rightarrow
\mathcal{G}_{q}[\alpha],
\end{equation}
where $\mathcal{G}_{q}[\alpha]$ is the set of functions
$\{be_q^{-\beta |\xi|^{\alpha}}, \, b>0, \, \beta >0 \},$ and
$$
q^{L}=\frac{3+Q\alpha}{1+ \alpha}, \, Q=2q-1 \,,
$$
i.e.,
$$
\frac{2}{q^L-1}=\frac{1+\alpha}{1+\alpha(q-1)} \,.
$$
The particular case $q=Q=1$ recovers $q^L=\frac{3+\alpha}{1+
\alpha}$, already known in the literature \cite{PratoTsallis1999}.
Denote $ \mathcal{Q}_1=\{(Q,\alpha): 1 \le Q < 3, \, \alpha = 2 \},$
$ \mathcal{Q}_2=\{(Q,\alpha): 1 \le  Q < 3, \, 0<\alpha < 2 \}$ and
$\mathcal{Q} = \mathcal{Q}_1 \cup \mathcal{Q}_2.$ Note that the case
$(Q,\alpha) \in \mathcal{Q}_1$ for $q$-independent random variables
with a finite $Q$-variance was studied in
\cite{UmarovTsallisSteinberg}. For $(Q,\alpha) \in \mathcal{Q}_2$
the $Q$-variance is infinite. We will focus our analysis namely on
the latter case. Note that the case $\alpha=2$, in the framework of
this classification like the classic $\alpha$-stable distributions,
becomes peculiar.

In the scope of second classification we study the attractors of
scaled sums, and expand the results of paper
\cite{UmarovTsallisSteinberg} to the region $\mathcal{Q}$
generalizing the mapping (\ref{paper1}) to
\begin{equation}
\label{paper2part2} F_{\zeta_{\alpha}(q)}: \mathcal{G}_q [\alpha]
\rightarrow \mathcal{G}_{z_{\alpha}(q)}[\alpha], \, 1 \le q<2, \,
0<\alpha \leq 2,
\end{equation}
where
\[
\zeta_{\alpha}(s) = \frac{\alpha - 2(1-q)}{\alpha} \, \, \mbox{and}
\, \, z_{\alpha}(s)=\frac{\alpha q + 1 - q}{\alpha + 1 - q}.
\]
Note that, if $\alpha = 2$, then $\zeta_{2}(q)=q$ and
$z_2(q)=(1+q)/(3-q),$ thus recovering the mapping (\ref{paper1}),
and consequently, the result of \cite{UmarovTsallisSteinberg}.

These two classifications of $(q,\alpha)$-stable distributions based
on mappings (\ref{paper2part1}) and (\ref{paper2part2})
respectively, can be unified to the scheme
\begin{equation}
\label{schemeLevy3} \mathcal{L}_q[\alpha]
\stackrel{F_{q}}{\longrightarrow} \mathcal{G}_q[\alpha]
\stackrel{F_{q_{\zeta(\alpha)}}}{\longleftrightarrow}
\mathcal{G}_{q_{\zeta(\alpha)}}[2]
\end{equation}
\hspace{1.5in} $\updownarrow \, F_{q}$
\[
\hspace{0.1in} \mathcal{G}_{q^{L}} [2],
\]
which gives the full picture of interrelations of
$(q,\alpha)$-stable distributions with parameters $q \in [1,2)$ and
$\alpha \in(0,2)$ (see details in Section VIII).

\section{Preliminaries and auxiliary results}
\subsection{Basic operations of $q$-algebra} In this section we
briefly recall the basic operations of $q$-algebra. Indeed, the
analysis we will conduct is entirely based on the $q$-structure of
nonextensive statistical mechanics (for more details see
\cite{GellmannTsallis,BoonTsallis,encyclopedia} and references
therein). To this end, we recall the well known fact that the
classical Boltzmann-Gibbs entropy $S_{BG}=-\sum_i p_i \ln p_i$
satisfies the additivity property. Namely, if $A$ and $B$ are two
independent subsystems, then $ \label{additivity}
S_{BG}(A+B)=S_{BG}(A)+S_{BG}(B). $ However, the $q$-generalization
of the classic entropy introduced in \cite{Tsallis1988} and given by
$S_q=\frac{1-\sum_i p_i^q}{q-1}$ with $q\in \cal{R}$ and $S_1 =
S_{BG}$, does not possess this property if $q \neq 1$. Instead, it
satisfies the {\it pseudo-additivity} (or {\it $q$-additivity})
\cite{Tsallis1988,PratoTsallis1999,BoonTsallis}
\[
S_q(A+B)=S_q(A)+S_q(B)+(1-q)\,S_q(A)\,S_q(B).
\]
Inherited from the right hand side of this equality, the {\it
$q$-sum} of two given real numbers, $x$ and $y$, is defined as $x
\oplus_q y = x+y+(1-q)xy$. The $q$-sum is commutative, associative,
recovers the usual summing operation if $q=1$ (i.e. $x \oplus_1 y =
x+y$), and preserves $0$ as the neutral element (i.e. $x \oplus_q 0
= x$). By inversion, we can define the {\it $q$-subtraction} as $x
\ominus_q y = \frac{x-y}{1+(1-q)y}.$ The {\it $q$-product} for $x,y$
is defined by the binary relation $x \otimes_q y =
[x^{1-q}+y^{1-q}-1]_+^{\frac{1}{1-q}}.$ Here the symbol $[x]_+$
means that $[x]_+ = x$ if $x \geq 0$, and $[x]_+ = 0$ if $x < 0.$
This operation also commutative, associative, recovers the usual
product when $q=1$, and preserves $1$ as the unity. The $q$-product
is defined if $x^{1-q}+y^{1-q} \geq 1$. Again by inversion, it can
be defined the {\it $q$-division}: $x \oslash_q y =
(x^{1-q}-y^{1-q}+1)^{\frac{1}{1-q}}.$

\subsection{$q$-generalization of the exponential and cyclic
functions}
\par
Now let us recall the main properties of two functions,
$q$-exponential and $q$-logarithm, which will be essentially used in
this paper. Let $e_q^x$ and $ln_q x$ denote respectively the
functions
$$e_q^x=[1+(1-q)x]_+^{\frac{1}{1-q}},\ \ {\rm and} \ \ln_q x=
\frac{x^{1-q}-1}{1-q}, \, (x>0).$$ The entropy $S_q$ then can be
conveniently rewritten in the form $S_q=\sum_i p_i \ln_q
\frac{1}{p_i}.$
For the $q$-exponential the relations $e_q^{x \oplus_q y} = e_q^x
e_q^y$ and $e_q^{x+y}=e_q^x \otimes_q e_q^y$ hold true. These
relations can be written equivalently as follows: $\ln_q (x
\otimes_q y)=\ln_q x + \ln_q y,$ and $\ln_q (x y)=(\ln_q x) \oplus_q
(\ln_q y)$. The $q$-exponential and $q$-logarithm have the
asymptotics
\begin{equation}
\label{exp} e_q^x = 1 + x + \frac{q}{2}x^2 + o(x^2), \, x
\rightarrow 0,
\end{equation}
and
\begin{equation}
\label{log} \ln_q (1+x) = x - \frac{q}{2} x^2 + o(x^2), \, x
\rightarrow 0,
\end{equation}
respectively. The $q$-product and $q$-exponential can be extended to
complex numbers $z=x+iy$ (see \cite{UmarovTsallisSteinberg,QT,QT2}).
In addition, for $q \neq 1$ the function $e_q^z$ can be analytically
extended to the complex plain except the point $z_0=-1/(1-q)$ and
defined as the principal value along the cut $(-\infty,z_0).$  If $q
< 1,$ then, for real $y$, $|e_q^{iy}| \geq 1$ and $|e_q^{iy}| \sim
K_q (1+y^2)^{\frac{1}{2(1-q)}}, \, y \rightarrow \infty,$ with
$K_q=(1-q)^{1/(1-q)}.$ Similarly, if $q > 1$, then $0 < |e_q^{iy}|
\leq 1$ and $|e_q^{iy}| \rightarrow 0$ if $|y| \rightarrow \infty.$
For complex $z$ it is not hard to verify the power series
representation
\begin{equation} \label{psr}
e_q^z = 1 +z + z^2 \sum_{n=0}^{\infty} \frac{A_n(q)}{(n+2)!} z^n, \,
\, |z|< \frac{1}{|1-q|},
\end{equation}
where $A_n(q)= \prod_{k=0}^{n} a_k(q),$ ~ $a_k(q)=q-k(1-q).$
%
Let $I_q = (-1/|1-q|, 1/|1-q|).$ Then it follows from (\ref{psr})
that for arbitrary real number $x\in I_q$ the equation
\begin{align*}
e_q^{ix} &= \{ 1 - x^2 \sum_{n=0}^{\infty} \frac{(-1)^n
A_{2n}(q)}{(2n+2)!} x^{2n} \}\\
         &+
               i \{  x - x^2 \sum_{n=0}^{\infty} \frac{(-1)^n
A_{2n+1}(q)}{(2n+3)!} x^{2n+1}  \}
\end{align*}
holds.
%
Define for $x \in I_q$ the functions $q$-cos and $q$-sin by formulas
\begin{equation}
\label{cos} \cos_q(x) = 1 - x^2 \sum_{n=0}^{\infty} \frac{(-1)^n
A_{2n}(q)}{(2n+2)!} x^{2n},
\end{equation}
and
\begin{equation}
\label{sin} \sin_q (x) = x - x^2 \sum_{n=0}^{\infty} \frac{(-1)^n
A_{2n+1}(q)}{(2n+3)!} x^{2n+1}.
\end{equation}
In fact, $\cos_q(x)$ and $\sin_q(x)$ can be defined for all real $x$
by using appropriate power series expansions. Properties of $q$-sin,
$q$-cos, and corresponding $q$-hyperbolic functions, were studied in
\cite{qborges1}. Here we note that the $q$-analogs of Euler's
formulas read
$$ e_q^{ix}=\cos_q(x)+i \, \, \sin_q(x), $$
{and} ~~

$$\cos_q(x)=\frac{e_q^{ix}+e_q^{-ix}}{2},  \quad
\sin_q(x)=\frac{e_q^{ix}-e_q^{-ix}}{2i}.$$
%
It follows from the definitions of $\cos_q (x)$ and $\sin_q (x),$
and from the equality $(e_q^{x})^2 = e_{(1+q)/2}^{2x}$ (see Lemma
2.1 in \cite{UmarovTsallisSteinberg}), that
\begin{equation}
\label{cos2x} \cos_q (2x) = e_{2q-1}^{2(1-q)x^2} -2 \, \sin_{2q-1}^2
(x).
\end{equation}
Denote $\Psi_q(x)=\cos_q 2x -1.$ Then equation (\ref{cos2x}) implies
\begin{equation}
\label{psi} \Psi_{q}(x)= ( e_{2q-1}^{2(1-q)x^2} - 1 ) -2 \,
\sin_{2q-1}^2 (x).
\end{equation}
The following two properties of $\Psi_{q}$ will be used later on.
\begin{proposition}
\label{lempsi} Let $q \ge 1$. Then
\begin{enumerate}
\item[1.]
$-2 \le \Psi_q(x) \le 0;$
\item[2.]
$\Psi_q(x) = - 2 \, q \, x^2 + o(x^3), \, x \rightarrow 0.$
\end{enumerate}
\end{proposition}
\vspace{.3cm}

{\it Proof.} Assume $q \le 1.$ Since $e_{2q-1}^{-2(q-1)x^2} \le 1,$
then (\ref{psi}) immediately implies that $\Psi_q(x) \le 0.$
Further, $\sin_{q}(x)$ can be written in the form (see
\cite{qborges1}) $\sin_q(x)=\rho_q(x) \, \sin[\varphi_q(x)],$ where
$\rho_q(x)=(e_q^{(1-q)x^2})^{1/2}$ and
$\varphi_q(x)=\frac{\arctan(1-q)x}{1-q}.$ A simple calculation
yields $\Psi_q (x) \ge -2.$ Using the asymptotic relation
(\ref{exp}), we get
\begin{equation}
\label{one} e_{2q-1}^{2(1-q)x^2} - 1 = 2(1-q)x^2 +o(x^3), \, x
\rightarrow 0.
\end{equation}
In turn, it follows from (\ref{sin}) that
\begin{equation}
\label{two} -2 \, \sin_{2q-1}^2 (x) = -2 \, x^2 + o(x^3),
x\rightarrow 0.
\end{equation}
Now (\ref{psi}), (\ref{one}) and (\ref{two}) imply the second part
of the statement. \eproof \vspace{.3cm}

Representation (\ref{psr}) shows the behaviour of $q$-exponential
near the origin.  It is not hard to verify that in the case $q>1$
for $x>(q-1)^{-1}$ the representation
\begin{align*}
e_q^{-x} &=      [(q-1)x]^{-\frac{1}{q-1}}
          \{1-\frac{1}{(1-q)^2 x}\\
         &+      \frac{1}{(1-q)^4 x^2} \sum_{n=0}^{\infty}\frac{(-1)^n
A_n(q)}{(n+2)!(q-1)^{2n}}(\frac{1}{x})^n\}
\end{align*}
holds.

\subsection{$q$-Fourier transform for symmetric densities} The {\it
$q$-Fourier transform} for $q \ge 1$ was introduced in
\cite{UmarovTsallisSteinberg} and used as a basic tool in
establishing the $q$-analog of the standard central limit theorem.
Formally the $q$-Fourier transform
 for a given function $f(x)$ is defined by
\begin{equation}
\label{FourierTr} F_q[f](\xi) = \int_{-\infty}^{\infty} e_q^{ix\xi}
\otimes_q f(x) dx \,.
\end{equation}
For discrete functions $f_k, k=0, \pm 1,...,$ this definition takes
the form
\begin{equation}
\label{FourierDiscrete} F_q[f](\xi) = \sum_{k= -\infty}^{\infty}
e_q^{ik\xi} \otimes_q f(k) \,.
\end{equation}
In the future we use the same notation in both cases. We also call
(\ref{FourierTr}) or (\ref{FourierDiscrete}) the {\it
$q$-characteristic function} of a given random variable $X$ with an
associated density $f(x),$ using the notations $F_q[X]$ or $F_q[f]$
equivalently.
\par
It should be noted that, if in the formal definition
(\ref{FourierTr}) $f$ is compactly supported then integration has to
be taken over this support, although, in contrast with the usual
analysis, the function $e_q^{ix\xi} \otimes_q f(x)$ under the
integral does not vanish outside the support of $f$. This is an
effect of the $q$-product.
\par
%
The $q$-Fourier transform for nonnegative $f(x)$ can be written in the
form
\begin{equation}
\label{identity2} F_q[f](\xi) = \int_{-\infty}^{\infty}f(x)
e_q^{{ix\xi}{(f(x))^{q-1}}}dx.
\end{equation}
%
We note that, if the $q$-Fourier transform of $f(x)$ defined by
(\ref{FourierTr}) exists, then it coincides with (\ref{identity2}).
The $q$-Fourier transform determined by the formula
(\ref{identity2}) has an advantage to compare to the formal
definition: it does not use the $q$-product, which is, as we noticed
above, restrictive in use (for $q \ge 1$).
\par
\begin{proposition}
\label{cos2} Let $f(x)$ be an even function. Then its $q$-Fourier
transform can be written in the form
\begin{equation}
\label{identity3} F_q[f](\xi) = \int_{-\infty}^{\infty}f(x) \cos_q
({x \xi [f(x)]^{q-1}})dx.
\end{equation}
\end{proposition}
\vspace{.3cm}

{\it Proof.} Notice that, because of the symmetry of $f$,
\[
\int_{-\infty}^{\infty}e_q^{ix\xi} \otimes_q f(x)dx =
\int_{-\infty}^{\infty}e_q^{-ix \xi} \otimes_q f(x)dx  \,.
\]
Taking this into account, we have
\[
F_q[f](\xi) = \frac{1}{2} \int_{-\infty}^{\infty} \left( e_q^{ix\xi}
\otimes_q f(x) +  e_q^{- ix\xi} \otimes_q f(x) \right) dx \,.
\]
Now due to (\ref{identity2}) we obtain
\[
F_q[f](\xi) = \int_{-\infty}^{\infty} f(x) \frac{ e_q^{ix\xi
[f(x)]^{q-1}}  +  e_q^{- ix\xi [f(x)]^{q-1} }}{2}  dx \,,
\]
which coincides with (\ref{identity3}). \eproof

Further, denote
\begin{align*} \label{hq}
H_{q,\alpha}=\{f \in L_1: f(x) \sim C
|x|^{-\frac{1+\alpha}{1+\alpha(q-1)}}, \, \, |x|\rightarrow
\infty\}.
\end{align*}
For a given $f \in H_{q,\alpha}$ the constant $C=C_f$ is defined
uniquely by $f.$  It is readily seen that $\phi
(q,\alpha)=\frac{\alpha+1}{1+\alpha(q-1)}>1$ for all $\alpha \in
(0,2)$ and $q \in [1,2).$ Moreover, $\phi(q,\alpha)(2q-1)<3$ for all
$\alpha \in (0,2)$ and $q \in [1,2),$ which implies
$\sigma_{2q-1}^2(f)=\infty.$ Notice also
$\phi(q,\alpha)=1+\alpha^{\ast}(q,\alpha),$ where
\[
\alpha^{\ast} = \alpha^{\ast}(q,\alpha)=\frac{\alpha(2-q)}{1+\alpha
(q-1)}.
\]
On the other hand, the density $g$ of any $\alpha^{\ast}$-stable
L\'evy distributions has the asymptotic behaviour $g(x) \sim
C/|x|^{1+\alpha^{\ast}}, ~ |x| \rightarrow \infty.$ Hence, for a
fixed $q \in [1,2)$ $H_{q,\alpha}$ is asymptotically equivalent to the set of
densities of $\alpha^{\ast}$-stable L\'evy distributions.
\par
The following proposition plays a key role in our further analysis.

\begin{proposition}
\label{mainlemma} Let $f(x), \, x \in R, $ be a symmetric
probability density function. Further, let either
\begin{itemize}
\item[(i)] the $(2q-1)$-variance $\sigma_{2q-1}^2 (f) <
\infty,$ (associated with $\alpha =2,$ and $1\le q<2$), or
\item[(ii)]
$f(x) \in H_{q,\alpha}$, where $(2q-1, \alpha) \in \mathcal{Q}_2$.
\end{itemize}

Then, for the $q$-Fourier transform of $f(x)$, the following
asymptotic relation holds true:
\begin{equation}
\label{asforfi} F_q[f](\xi)= 1 - \mu_{q,\alpha} |\xi|^{\alpha} +
o(|\xi|^{\alpha}), \xi \rightarrow 0,
\end{equation}
where
\begin{equation}
\mu_{q,\alpha} = \left\{ \begin{array}{ll}
  \vspace{1cm}
          \frac{q}{2} \sigma_{2q-1}^2 \nu_{2q-1} , &
          \mbox{if $\alpha = 2$ \,;} \\
          \   \frac{2^{2-\alpha}(1+\alpha(q-1))C_f}{2-q}  \int_0^{\infty}
\frac{- \, \Psi_q (y)}{y^{\alpha+1}} dy, & \mbox{if $(2q-1, \alpha)
\in \mathcal{Q}_2$ \,.}
  \end{array} \right.
\end{equation}
with $\nu_{2q-1}(f)= \int_{-\infty}^{\infty} [f(x)]^{2q-1}  \, dx$ .
\end{proposition}
%

{\it Proof.} First, assume that $\alpha = 2$.  By Proposition
\ref{cos2}
\begin{align}
\label{step_10} F_q [f](\xi) &= \int_{-\infty}^{\infty}(e_q^{ix
\xi}) \otimes_q f(x) dx \notag\\    &=
                              \int_{-\infty}^{\infty} f(x)
\cos_q (x \xi [f(x)]^{q-1}) dx  \,.
\end{align}
Making use of the asymptotic expansion (\ref{exp}) we can rewrite
the right hand side of (\ref{step_10}) in the form
\begin{align}
F_q[f](\xi) &=
   \int_{-\infty}^{\infty} f(x) \{1 + i x \xi [f(x)]^{q-1}\notag\\
&- q/2 x^2 \xi^2 [f(x)]^{2(q-1)} \} dx +    o(\xi^3 ) \notag\\
&=  1 - (q/2) \xi^2 \sigma^{2}_{2q-1} \nu_{2q-1}+ o(\xi^3 ), \, \xi
\rightarrow 0,
\end{align}
from which the first part of Proposition follows.

Now, we assume $(2q-1, \alpha) \in \mathcal{Q}_2.$ We apply
Proposition \ref{cos2} to obtain
\begin{align*}
F_q[f](\xi)-1 &= \int_{-\infty}^{\infty} f(x) [\cos_q ({x \xi
[f(x)]^{q-1}}) - 1] dx \\
&= 2 \int_{0}^{N} f(x) \Psi_q ( \frac {x \xi [f(x)]^{q-1}}{2})dx\\
&+ 2 \int_{N}^{\infty} f(x) \Psi_q (\frac {x \xi [f(x)]^{q-1}}{2})
dx \,,
\end{align*}
where $N$ is a sufficiently large finite number. In the first
integral we use the asymptotic relation $\Psi({x \over 2})= - {q
\over 2} x^2 + o(x^3), ~ x \rightarrow 0$, which follows from
Proposition \ref{lempsi}, and get
\par
$ 2 \int_{0}^{N} f(x) \Psi_q (\frac {x \xi [f(x)]^{q-1}}{2})dx = $
\begin{equation}
\label{1integral} -q \xi^2 \int_0^N x^2 f^{2q-1}(x)dx + o(\xi^3), \,
\xi \rightarrow 0,
\end{equation}
that is a quantity of order $o(|\xi|^{\delta}), \xi \rightarrow 0,$
for any $\delta < 2.$ In the second integral taking into account the
hypothesis of the proposition with respect to $f(x)$, we have
\begin{align}
&2 \int_{N}^{\infty} f(x) \Psi_q ( \frac {x \xi [f(x)]^{q-1}}{2}) dx \\
&= 2C_f \int_{N}^{\infty} \frac{1}{x^{\frac{\alpha +
1}{1+\alpha(1-q)}}} \Psi_q ( \frac {x^{1-\frac{(\alpha
+1)(q-1)}{1+\alpha(q-1)}} \xi}{2 C_f^{1-q}} ) dx \,.
\end{align}
We use the substitution
\[
x^{\frac{2-q}{1+\alpha(q-1)}} = \frac{2y}{C_f^{q-1}\xi}
\]
in the last integral, and obtain
\begin{align}
2 \int_{N}^{\infty} &f(x) \Psi_q (\frac{ x \xi [f(x)]^{q-1}}{2}) dx
\notag \\
&= \mu_{q,\alpha}|\xi|^{\alpha}+o(|\xi|^{\alpha}), ~ \xi \to 0,
\label{2integral}
\end{align}
where
\begin{equation*}
 \mu_{q,\alpha} = -\frac{2^{2-\alpha}(1+\alpha(q-1))C_f}{2-q}
 \, \, \int_0^{\infty} \frac{\Psi_q (y)}{y^{\alpha+1}} dy.
 \end{equation*}
Hence, the obtained asymptotic relations (\ref{1integral}) (we take
$\delta \in (\alpha,2)$) and (\ref{2integral}) complete the proof.
\eproof

For stable distributions $\mu_{q, \alpha}$ must be positive. We have
seen (Proposition \ref{lempsi}) that if $q\ge1$, then $\Psi_q(x) \le
0$ (not being identically zero), which yields $\mu_{q,\alpha} > 0
\,.$ Note also that the condition for $f(x)$ to be symmetric was not
required in \cite{UmarovTsallisSteinberg}, if $\sigma_{2q-1} (f) <
\infty.$

\section{Weak convergence of correlated random variables}

Let us start this section by introducing the notion of {\it
$q$-independence}. We will also introduce two types of convergence,
namely, {\it $q$-convergence} and {\it weak $q$-convergence} and
establish their equivalence.

By definition, two random variables $X$ and $Y$ are said to be
\textit{$(q^{'},q,q^{''})$-independent}
 if
\begin{equation}
\label{q-correlation} F_{q^{'}}[X+Y](\xi)=F_q[X](\xi)
\otimes_{q^{''}} F_q[Y](\xi) \,.
\end{equation}
In terms of densities, equation (\ref{q-correlation}) can be
rewritten as follows. Let $f_X$ and $f_Y$ be densities of $X$ and
$Y$ respectively, and let $f_{X+Y}$ be the density of $X+Y$. Then
\begin{equation}
\label{q-correlation2} \int_{-\infty}^{\infty} e_{q^{'}}^{i x \xi}
\otimes_{q^{'}} f_{X+Y}(x) dx = F_q[f_X](\xi) \otimes_{q^{''}}
F_q[f_Y](\xi).
\end{equation}
If all three parameters $q^{'}, q$ and $q^{''}$ coincide, i.e.
$q=q^{'}=q^{''},$ then we call simply $q$-independent. For $q=1$ the
condition (\ref{q-correlation}) turns into the well known relation
\[
F[f_X \ast f_Y] = F[f_X] \cdot F[f_Y]
\]
between the convolution (noted $\ast$) of two densities and the
multiplication of their (classical) characteristic functions, and
holds for independent $X$ and $Y$. If $q \neq 1$, then
$(q^{'},q,q^{''})$-independence describes a specific class of
correlations.

\begin{remark}
\begin{em}
{It is worth to mention at this point that $q$-independence
appears to be relevant to the notion of \textit{scale-invariance}
\cite{RodriguezSchwammleTsallis}. To be more specific, it might well
be that $q$-independence implies scale-invariance, i.e.,
scale-invariance is necessary for $q$-independence, although it is
by now clear that it is not sufficient. Indeed, scale-invariant
probabilistic models exist in the literature. Some of them
presumably involve $q$-independence since their $N\to\infty$ limits
are $q$-Gaussians
\cite{RodriguezSchwammleTsallis,HanelThurnerTsallis}; others do not
involve $q$-independence
\cite{MoyanoTsallisGellMann,ThistletonMarshNelsonTsallis,HilhorstSchehr}
(if they did involve, their $N\to\infty$ limits would have to be
$q$-Gaussians and they are not)}. See also paper
\cite{HahnJiangUmarov} which discusses limit distributions in
general setting within the exchangeability concept.
\end{em}
\end{remark}

%
Let $X_N$ be a sequence of identically distributed random variables.
Denote $Y_N=X_1+...+X_N$. By definition, \textit{ $X_N$ is said to
be $(q^{'},q,q^{''})$-independent (or $(q^{'},q,q^{''})$-i.i.d.)} if
the relations
\begin{equation}
\label{qiid1} F_{q^{'}}[Y_N ](\xi)=F_{q}[X_1](\xi) \otimes_{q^{''}}
... \otimes_{q^{''}} F_{q}[X_N](\xi)
\end{equation}
hold for all $N=2,3,...$.

For $q=q^{'}=q^{''}=1$ the condition (\ref{qiid1}) turns into the
condition for the sequence $X_N$ to be usual i.i.d. If
$q=q^{'}=q^{''}$ then we call the sequence $X_N$ simply a $q$-i.i.d.
Consider example of an $(q^{'},q,q)$-i.i.d. sequence of random
variables, where $q \in (1,3)$ and $q^{'}=(3q-1)/(q+1).$ Assume
$X_N$ is the sequence of identically distributed random variables
with the associated Gaussian density
\[
G_{q^{'}}(\beta, x)=\frac{\sqrt{\beta}}{C_{q^{'}} } e_{q^{'}}^{-
\beta x^2},
\] where
$C_{q^{'}}$ is the normalizing constant (see, e.g.
\cite{UmarovTsallisSteinberg}). Further, assume the sums
$X_1+...+X_N, \, N=2,3,...,$ are distributed according to the
density $G_{q^{'}}(\alpha, x)$, where $\alpha =
N^{-\frac{1}{2-q^{'}}} \beta$. Then the sequence $X_N$ satisfies
(\ref{qiid1}) for all $N=2,3,...$, with $q=q^{''},$ thus being
$(q^{'},q,q)$-independent identically distributed sequence of random
variables.

For the sake of simplicity in this paper we will consider only
$q$-i.i.d. random variables.

%
By definition, a sequence of random variables $X_N$ is said to be
\textit{$q$-convergent} to a random variable $X_{\infty}$ if
$\lim_{N \rightarrow \infty} F_q [X_N](\xi) = F_q [X_{\infty}](\xi)$
locally uniformly in $\xi$.

Evidently, this definition is equivalent to the weak convergence
(denoted by "$\Rightarrow$") of random variables if $q=1.$ For $q
\neq 1$ denote by $W_q$ the set of continuous functions $\phi$
satisfying the condition $|\phi(x)| \leq C(1+|x|)^{-\frac{q}{q-1}},
\, x \in R$.

A sequence of random variables $X_N$ with the density $f_N$ is
called \textit{weakly $q$-convergent} to a random variable
$X_{\infty}$ with the density $f$ if $\int_{R} f_N(x) d m_q
\rightarrow \int_{R^d} f(x)d m_q $ for arbitrary measure $m_q$
defined as $dm_q(x)= \phi_q (x) dx,$ where $\phi_q \in W_q$. We
denote the $q$-convergence by the symbol
$\stackrel{q}{\Rightarrow}$.

\begin{proposition}
\label{weak} Let $q>1.$ Then $X_N \Rightarrow X_0 $ yields $X_N
\stackrel{q}{\Rightarrow} X_0$.
\end{proposition}

The proof of this statement immediately follows from the obvious
fact that $W_q$ is a subset of the set of bounded continuous
functions. Recall that a sequence of probability measures $\mu_N$ is
called {\it tight} if, for an arbitrary $\epsilon >0$, there is a compact
$K_{\epsilon}$ and an integer $N^{\ast}_{\epsilon}$ such that
$\mu_N(R^d \setminus K_{\epsilon}) < \epsilon$ for all $N \geq
N_{\epsilon}^{\ast}$.

\begin{proposition}
\label{tight} Let $1<q<2.$ Assume a sequence of random variables
$X_N$, defined on a probability space with a probability measure
$P$, and associated densities $f_N,$ is $q$-convergent to a random
variable $X$ with an associated density $f.$ Then the sequence of
associated probability measures $\mu_N = P(X_N^{-1})$ is tight.
\end{proposition}

{\it Proof.} Assume that $1<q<2$ and $X_N$ is a $q$-convergent
sequence of random variables with associated densities $f_N$ and
associated probability measures $\mu_N$. We have
\[
\frac{1}{R}\int_{-R}^{R}(1-F_q[f_N](\xi)) d \xi =
\frac{1}{R}\int_{-R}^{R}(1-\int_{R} f_N e_q^{i x \xi f_N^{q-1}} dx )
d \xi
\]
\begin{equation}
\label{tight1} =\int_{R} \left( \frac{1}{R}\int_{-R}^{R}(1- e_q^{i x
\xi f_N^{q-1}} ) d \xi \right) d \mu_N(x).
\end{equation}
It is not hard to verify that
\begin{equation}
\label{tight2} \frac{1}{R}\int_{-R}^{R} e_q^{i x \xi t} d \xi =
\frac{2 \sin_{\frac{1}{2-q}} (R x(2-q) t)}{Rx(2-q)t}.
\end{equation}
It follows from (\ref{tight1}) and (\ref{tight2}) that
\begin{align}
 \frac{1}{R}&\int_{-R}^{R}(1-F_q[f_N](\xi)) d \xi
\notag \\ & 2\int_{-\infty}^{\infty} \left( 1 - \frac{\sin_{
\frac{1}{2-q} } (x(2-q)R f_N^{q-1})}{Rx(2-q)f_N^{q-1}} \right)
d\mu_N(x).
\label{tight3}
\end{align}
Since $1<q<2$ by assumption,  $\frac{1}{2-q} > 1$ as well. It is
known \cite{qborges1,qnivanen,qborges2} that for any $q^{'}>1$ the
properties $sin_{q^{'}}(x) \leq 1$ and $(sin_{q^{'}}(x))/x
\rightarrow 1, \, x \rightarrow 0$ hold. Moreover,
$(\sin_{q^{'}}(x))/x \leq 1, \forall x \in R.$ Suppose,
$lim_{|x|\rightarrow \infty} |x|f_N^{q-1}=L_N, \, N \ge 1.$ Divide
the set $\{N \ge N_0\}$ into two subsets $A=\{N_j \ge N_0: L_{N_j} >
1\}$ and $B=\{N_k \ge N_0: L_{N_k} \le 1\}.$
If $N \in A,$ since $\sin_{\frac{1}{2-q}} \le 1,$ there is a number
$a>0$ such that
\begin{align*}
\frac{1}{R}\int_{-R}^{R}(1 & - F_q[f_N](\xi)) d \xi \\ & \geq 2
\int_{|x| \geq a} \left ( 1-\frac{1}{R|x|(2-q)f_N^{q-1}} \right)
d\mu_N (x) \\
&\geq C \mu_N \left( |x| \geq a \right), \, \, C>0 \, ~~ \forall \,
N \in A,
\end{align*}
for $R$ small enough. Now taking into account the $q$-convergence of
$X_N$ to $X$ and, if necessary, taking $R$ smaller, for any
$\epsilon
>0$, we obtain
\[
\mu_N \left( |x| \geq a \right) \leq \frac{1}{C
R}\int_{-R}^{R}(1-F_q[f_0](\xi)) d \xi < \epsilon, \,  \, ~~ \forall
\, N \in A.
\]
If $N \in B,$ then there exist constants $b>0, \, \delta>0,$ such
that
\[
f_N(x) \le \frac{L_N+\delta}{|x|^{\frac{1}{q-1}}} \le
\frac{1+\delta}{|x|^{\frac{1}{q-1}}}, \, |x|\ge b, \forall \, N \in
B.
\]
Hence, we have
\begin{align*}
\mu_N(|x|>b) &= \int_{|x|>b}f_N(x)dx \\& \le {(1+\delta)}
\int_{|x|>b}\frac{dx}{|x|^{\frac{1}{q-1}}}, \, N \in B.
\end{align*}
Since, $1/(q-1)>1,$ for any $\epsilon >0$ we can select a number
$b_{\epsilon} \ge b$ such that $\mu_N(|x|>b_{\epsilon})<\epsilon, \,
N \in B.$ As far as $A \cup B = \{N \ge N_0\}$ the proof of the
statement is complete. \eproof

Further, we introduce  the function
\begin{align}
 D_q(t)= D_q(t; a)&=t e_q^{i a t^{q-1}} \notag\\
\label{dt}
&= t(1 + i (1-q) a t^{q-1})^{-\frac{1}{q-1}},
\end{align}
defined on  $ [0,1]$, where $1<q<2$ and $a$ is a fixed real number.
Obviously, $D_q(t)$ is continuous on $[0,1]$ and differentiable in
the interval $(0,1)$. In accordance with the classical Lagrange
average theorem for any $t_1, t_2, \, \, 0 \leq t_1 < t_2 \leq 1$
there exists a number $t_{\ast}, \, \, t_1 < t_{\ast} < t_2$ such
that
\begin{equation}
\label{ev} D_q(t_1)-D_q(t_2) = D_q^{'}(t_{\ast}) (t_1 - t_2),
\end{equation}
where $D_q^{'}$ means the derivative of $D_q(t)$ with respect to
$t$.

Consider the following Cauchy problem for the Bernoulli equation
\begin{equation}
\label{bern1} y^{'} - \frac{1}{t} y = \frac{ia(q-1)}{t} y^q, \, \,
y(0)=0,
\end{equation}
It is not hard to verify that $y(t)=D_q(t)$ is a solution to problem
(\ref{bern1}).

\begin{proposition}
For $D^{'}_q (t)$  the estimate
\begin{equation}
\label{est} |D_q^{'}(t;a)| \leq C (1 + |a|)^{-\frac{q}{q-1}}, \, \,
t \in (0,1], \, a \in R^1,
\end{equation}
holds, where constant $C$ does not depend on $t$.
\end{proposition}

{\it Proof.} It follows from (\ref{dt}) and (\ref{bern1}) that
\begin{align*}
|y^{'}(t)| &\leq t^{-1}|y + i a (q-1)y^{q}| \\& = |e_q^{i a t^{q-1}}
+ i a (q-1)t^{q-1}(e_q^{i a t^{q-1}})^q|\\&=
|1+ia(1-q)t^{q-1}|^{-\frac{q}{q-1}}  \\ &\leq C
(1+|a|)^{-\frac{q}{q-1}}, \, t \in (0,1].
\end{align*}
\eproof
\par
Now we are in a position to formulate the following two theorems on
the relationship between $q$-convergence and weak $q$-convergence.
\begin{theorem}
\label{continuitytheorem1} Let $1<q<2$ and a sequence of random
vectors $X_N$ be weakly $q$-convergent to a random vector $X$. Then
$X_N$ is $q$-convergent to $X$.
\end{theorem}

{\it Proof.} Assume $X_N,$ with associated densities $f_N,$ is
weakly $q$-convergent to a $X,$ with an associated density $f$. The
difference $\mathcal{F}_q [f_N](\xi) - \mathcal{F}_q [f](\xi)$ can
be written in the form
\begin{align}
 \mathcal{F}_q [f_N](\xi) &- \mathcal{F}_q
[f](\xi) \notag \\
\label{difference}       &=  \int_{R^d} \left( D_q(f_N(x)) - D_q(
f(x) ) \right) dx,
\end{align}
where $D_q(t)=D_q(t;a)$ is defined in (\ref{dt}) with $a = x \xi$.
It follows from (\ref{ev}) and (\ref{est}) that
\begin{align*}
|\mathcal{F}_q & [f_N](\xi) - \mathcal{F}_q [f](\xi)|\\ &\leq C
\int_{R^d} | (1+|x|)^{-\frac{q}{q-1}} \left(f_N(x)- f(x) \right)|dx,
\end{align*}
which yields $\mathcal{F}_q [f_N](\xi) \rightarrow \mathcal{F}_q
[f_N](\xi)$ for all $\xi \in R^d$. \eproof
\begin{theorem}
Let $1<q<2$ and a sequence of random vectors $X_N$ with the
associated densities $f_N$ is $q$-convergent to a random vector $X$
with the associated density $f$ and $\mathcal{F}_q[f](\xi)$ is
continuous at $\xi = 0$. Then $X_N$ weakly $q$-converges to $X$.
\end{theorem}

 {\it Proof.} Suppose that $f_N$ converges to $f$ in the sense of
$q$-convergence. It follows from Proposition \ref{tight} that the
corresponding sequence of induced probability measures $\mu_N =
P(X_N^{-1})$ is tight. This yields relatively weak compactness of
$\mu_N.$ Theorem \ref{continuitytheorem1} implies that each weakly
convergent subsequence $\{\mu_{N_j}\}$ of $\mu_N$ converges to
$\mu=P(X^{-1}).$ Hence, $\mu_N \Rightarrow \mu$, or the same, $X_N
\Rightarrow X.$ Now applying Proposition \ref{weak} we complete the
proof. \eproof

\section{Symmetric $(q,\alpha)$-stable distributions and their properties}

In this section we introduce the symmetric $(q,\alpha)$-stable
distributions and classify them on the base of mapping
(\ref{paper2part1}). In this classification $q$ takes any value in
$[1,2),$ however we distinguish the cases $\alpha = 2$ and $0<
\alpha < 2.$
\begin{definition}
\label{stabledistr} A random variable $X$ is said to have a
$(q,\alpha)$-stable distribution if its $q$-Fourier transform is
represented in the form $e_q^{-\beta |\xi|^{\alpha}}$, with $\beta
>0.$ We denote  the set of random variables with
$(q,\alpha)$-stable distributions by $\mathcal{L}_q[\alpha].$
\end{definition}
Denote $ \mathcal{G}_q[\alpha] = \{b \, e_q^{-\beta |\xi|^{\alpha}},
\, \, b > 0, \, \, \beta >0\}.$ In other words $X \in
\mathcal{L}_q[\alpha],$ if $F_q[X] \in \mathcal{G}_q[\alpha]$ with
$b=1.$ Note that if $\alpha = 2$, then $\mathcal{G}_q[2]$ represents
the set of $q$-Gaussians and $\mathcal{L}_q[2]$ - the set of random
variables whose densities are $q_{\ast}$-Gaussians, where
$q_{\ast}=(3q-1)/(1+q)$. Further, from the asymptotic relation
(\ref{exp}) we have $e_q^{-\beta|\xi|^{\alpha}}=1-\beta
|\xi|^{\alpha} + o(|\xi|^{\alpha}).$ This and Proposition
\ref{mainlemma} imply that the associated density of any
$(q,\alpha)$-stable distribution belongs to $H_{q,\alpha}.$
\begin{proposition}
\label{q-stability} Let $q$-independent random variables $X_j \in
\mathcal{L}_q[\alpha], j=1,..,m.$ Then for constants $a_1,...,a_m,$
\[
\sum_{j=1}^m a_j X_j \in \mathcal{L}_q[\alpha].
\]
\end{proposition}

{\it Proof.} Let $$F_q[X_j](\xi)=e_q^{-\beta_j}|\xi|^{\alpha}, \,
j=1,...,m.$$ Using the properties $e_q^x \otimes_q e_q^y =e_q^{x+y}$
and $F_q[aX](\xi)=F_q[X](a^{2-q}\xi),$ it follows from the
definition of the $q$-independence that
\[
F_q[\sum_{j=1}^m a_j X_j] = e_q^{-\beta |\xi|^{\alpha}}, \, \beta =
\sum_{j=1}^m \beta_j |a|^{\alpha(2-q)} > 0.
\]
\eproof

Proposition \ref{q-stability} justifies the stability of
distributions in $\mathcal{L}_q[\alpha].$ Recall that if $q=1$ then
$q$-independent random variables are independent in the usual sense.
Thus, if $q=1, \, 0<\alpha<2,$ then $\mathcal{L}_1[\alpha] \equiv
\mathcal{L}_{sym}[\alpha]$, where $\mathcal{L}_{sym}[\alpha]$ is the
set of $\alpha$-stable L\'evy distributions.
%

Moreover, the appropriately scaling limit of sequences of
$q$-independent random variables with $(q,\alpha)$-stable
distributions has again a $(q,\alpha)$-stable distribution. To this
end consider the sum
\[
Z_N = \frac{1}{s_{N}(q,\alpha)} \, (X_1 + ...+ X_N ), N=1,2,...
\]
 where $s_{N}(q,\alpha)$
is a scaling parameter specified below. First we prove a general
result.

\begin{theorem} \label{th3}Assume $(2q-1,\alpha) \in \mathcal{Q}_2.$ Let
$X_N$ be symmetric $q$-independent random variables all having the
same probability density function $f(x) \in H_{q,\alpha}.$ Then
$Z_N$, with $s_N(q,\alpha) = (\mu_{q,\alpha}N)^{\frac{1}{\alpha
(2-q)}}$, is $q$-convergent to a $(q,\alpha)$-stable distribution,
as $N \rightarrow \infty.$
\end{theorem}
\par
%
{\it Proof.} Assume $(Q,\alpha) \in \mathcal{Q}_2.$ Let $f$ be the
density associated with $X_1$. First we evaluate
$F_q[X_1]=F_q[f(x)].$ Using Proposition \ref{mainlemma} we have
\begin{equation}
\label{step_1} F_q[f](\xi)= 1 - \mu_{q,\alpha} |\xi|^{\alpha} +
o(|\xi|^{\alpha}), \xi \rightarrow 0.
\end{equation}
Denote $Y_j = N^{-\frac{1}{\alpha}} \, X_j, j=1,2,...$. Then $Z_N =
Y_1 +...+Y_N.$ Further, it is readily seen that for a given random
variable $X$ and real $a>0$, the equality $F_q
[aX](\xi)=F_q[X](a^{2-q} \xi)$ holds. It follows from this relation
that $F_q[Y_j]=F_q[f]( \frac{\xi}{ (\mu_{q,\alpha}N)^{1/\alpha } }
), \, j=1,2,...$ Moreover, it follows from the $q$-independence of
$X_1,X_2,...,$ and the associativity of the $q$-product that
\begin{align}
&F_q[Z_N](\xi)\notag
\\
\label{step2} &= F_q[f](
\frac{\xi}{(\mu_{q,\alpha}N)^{\frac{1}{\alpha} }})
\underbrace{\otimes_q ... \otimes_q}_{N \, \mbox{factors}} F_q[f](
(\frac{\xi}{(\mu_{q,\alpha}N)^{\frac{1}{\alpha} }}).
\end{align}
Further, making use of the expansion (\ref{log}) for the
$q$-logarithm, equation (\ref{step2}) implies
\begin{align}
\ln_q &F_q[Z_N](\xi)= N \ln_q F_q[f](
(\mu_{q,\alpha}N)^{-\frac{1}{\alpha}}\xi ) \notag \\ &= N \ln_q ( 1-
\frac{|\xi|^{\alpha}}{N} + o(\frac{|\xi|^{\alpha}}{N})) \notag \\ &=
\label{step101} - |\xi|^{\alpha} + o(1), \, N \rightarrow \infty,
\end{align}
locally uniformly by $\xi$. Hence, locally uniformly by $\xi,$
\begin{equation}
\label{step_3} \lim_{N \rightarrow \infty} F_q[Z_N] = e_q^{-
|\xi|^{\alpha}} \in \mathcal{G}_q[\alpha].
\end{equation}
Thus, $Z_N$ is $q$-convergent to a random variable with
$(q,\alpha)$-stable distribution, as $N \rightarrow \infty.$ \eproof

Since the density of $X \in \mathcal{L}_q[\alpha]$ is in
$H_q[\alpha]$ it follows immediately the following Corollary from
Theorem \ref{th3}.

\begin{corollary}
Assume $(2q-1,\alpha) \in \mathcal{Q}_2.$ Let $X_N$ be a sequence of
symmetric $q$-independent $(q,\alpha)$-stable random variables. Then
$Z_N$, with the same $s_N(q,\alpha)$ in Theorem \ref{th3},
$q$-weakly converges to a $(q,\alpha)$-stable distribution.
\end{corollary}

Note that $\alpha=2$ is not included to $\mathcal{Q}_2$ in Theorem
\ref{th3}. The case $\alpha=2$, in accordance with the first part of
Proposition \ref{mainlemma}, coincides with Theorem 2 of
\cite{UmarovTsallisSteinberg}. Recall that in this case
$\mathcal{L}_q[2]$ consists of random variables whose densities are
in $\mathcal{G}_{q^{\ast}}[2]$, where $q^{\ast}=\frac{3q-1}{q+1}.$

Theorem \ref{th3} also allows to establish a connection between the
classic L\'evy distributions and $q^{L}_{\alpha}$-Gaussians. Indeed,
for a $X\in \mathcal{L}_q[\alpha]$, its density function $f$ has
asymptotics
\[
f \sim C_f/  x^{ (\alpha +1)/(1 + \alpha (q-1)) }, \, \, |x|
\rightarrow \infty.
\]
It is not hard to verify that there exists a
$q^{L}_{\alpha}$-Gaussian, which is asymptotically equivalent to
$f$. Let us now find $q^{L}_{\alpha}.$ Any $q^{L}_{\alpha}$-Gaussian
behaves asymptotically $C_1/|x|^{\eta} = C_2/ |x|^{
2/(q^{L}_{\alpha}-1) }, \, C_j=const, \, j=1,2$, i.e.
$\eta=2/(q^{L}_{\alpha}-1).$ Hence, we obtain the relation
\begin{equation}\label{37}
\frac{\alpha +1}{1 + \alpha (q-1) } = \frac{2}{q^{L}_{\alpha}-1}.
\end{equation}
Solving this equation with respect to $q_{\alpha}^L$, we have
\begin{equation}
\label{general} q^{L}_{\alpha} = \frac{3+ Q \alpha }{\alpha + 1}, \,
\, Q=2q-1 \,,
\end{equation}
linking three parameters: $\alpha,$ the parameter of the
$\alpha$-stable L\'evy distributions,  $q,$ the parameter of
correlation, and $q^{L}_{\alpha}$, the parameter of attractors in
terms of $q^{L}_{\alpha}$-Gaussians. Equation (\ref{general})
identifies all $(Q,\alpha)$-stable distributions with the same index
of attractor $G_{q^{L}_{\alpha}},$ proving the following
proposition.

\begin{proposition}
\label{clas} Let $1\le Q <3 \, (Q=2q-1),$ $0<\alpha<2,$ and
\begin{align} \label{ql}
\frac{3+Q \alpha}{\alpha+1}=q_{\alpha}^L,
\end{align}
Then the density of $X \in \mathcal{L}_q[\alpha]$ is asymptotically
equivalent to $q_{\alpha}^L$-Gaussian.
\end{proposition}

In the particular case $Q=1$, we recover the known connection
between the classical L\'evy distributions ($q = Q = 1$) and
corresponding $q^{L}_{\alpha}$-Gaussians. In fact, putting $Q=1$ in
equation (\ref{general}), we obtain
\begin{equation}
\label{Q=1} q^{L}_{\alpha} = \frac{3+\alpha}{1+\alpha}, \, \,
0<\alpha<2.
\end{equation}
When $\alpha$ increases between $0$ and $2$ (i.e. $0<\alpha<2$),
$q^{L}_{\alpha}$ decreases between $3$ and $5/3$ (i.e. $5/3 <
q^L_{\alpha}<3$).

It is useful to find the relationship between $\eta =
\frac{2}{q^{L}_{\alpha}-1}$, which corresponds to the asymptotic
behaviour of the attractor depending on $(\alpha,Q)$. Using formula
(\ref{37}), we obtain
\begin{equation}
\label{eta} \eta = \frac{2(\alpha + 1)}{2 + \alpha (Q-1)}.
\end{equation}

\begin{proposition}
Let $X \in \mathcal{L}_Q[\alpha], \, 1\le Q <3, \, 0<\alpha<2.$ Then
the associated density function $f_X$ has asymptotics $f_X(x)\sim
|x|^{\eta}, \, |x|\rightarrow \infty,$ where $\eta=\eta(Q,\alpha)$
is defined in (\ref{eta}).
\end{proposition}

If $Q=1$ (classic L\'evy distributions), then (\ref{eta}) implies
the well-known fact $\eta = \alpha + 1.$

Analogous relationships can be obtained  for other values of $Q$. We
call, for convenience, a $(Q,\alpha)$-stable distribution  a
$Q$-Cauchy distribution, if $\alpha = 1.$ We obtain the classic
Cauchy-Poisson distribution if  $Q = 1$.
For $Q$-Cauchy distributions (\ref{ql}) and (\ref{eta}) imply
\begin{equation}
\label{Q-Cauchy} q^{L}_{1}(Q) = \frac{3+Q}{2} \, \, \, \mbox{and} \,
\, \, \eta=\frac{4}{Q+1},
\end{equation}
respectively.

\section{Scaling limits of sums of $(q,\alpha)$-stable distributions}

In this section we generalize the $q$-central limit theorem
established in \cite{UmarovTsallisSteinberg} for $q$-Gaussians, that
is in the case of $\alpha=2,$ to symmetrical $(q,\alpha)$-stables
with any $\alpha \in (0,2]$.

Let $1 < q < 2$, and $f \in \mathcal{G}_q [\alpha], \, \, 0<\alpha
\le 2.$ It follows from the definition of the $q$-exponential that
$f \sim C_f \, |x|^{-\alpha \over {q-1}}, \, C_f>0,$ as $|x| \to
\infty$. Analogously, if $g \in \mathcal{G}_q [2],$ then $g \sim C_g
\, |x|^{-2 \over {q-1}}, \, C_g>0,$ as $|x| \to \infty$. Comparing
orders of asymptotics we can easily verify that for a fixed $\alpha
\in (0,2]$ and for any $q \in (1,2)$ there exists a one-to-one
mapping
\[
\mathcal{M}_{q,q^{\ast}}: \mathcal{G}_{q} [\alpha] \rightarrow
\mathcal{G}_{q^{\ast}} [2], \quad q^{\ast} = \frac{\alpha +
2(q-1)}{\alpha},
\]
such that the image of a density $f \in \mathcal{G}_q [\alpha]$ is
again density. Analogously, there is a one-to-one mapping
\[
\mathcal{K}_{q,q^{\ast}}: \mathcal{G}_{q} [\alpha] \rightarrow
\mathcal{G}_{q^{\ast}} [2],
\]
with the same $q^{\ast},$ such that it maps $f(x)=e_q^{-\beta
|x|^{\alpha}},$ an element of $\mathcal{G}_q[\alpha]$ with the
coefficient $b=1$  onto the element
$g(x)=e_{q^{\ast}}^{-\frac{\alpha \beta}{2} |x|^2}$ with the same
coefficient $b=1.$ We notice that if $\alpha = 2,$ then $q^{\ast}=q$
and both operators coincide with the identity operator.

Let  $\mathcal{F}_q$ be an operator defined as
$\mathcal{F}_q=\mathcal{K}^{-1}_{z(q^{\ast}),q_{\ast}} F_{q^{\ast}}
\mathcal{M}_{q,q^{\ast}},$ where
$z(q^{\ast})=\frac{1+q^{\ast}}{3-q^{\ast}}.$
It is readily seen that in the particular case $\alpha=2$ it
coincides with the $q$-Fourier transform, $\mathcal{F}_q=F_q.$ We
call $\mathcal{F}_q$ a generalized $q$-Fourier transform.

\begin{proposition}
\label{formula1} Assume $0<\alpha \leq 2$ and let the numbers
$q^{\ast}, \, \, q_{\ast}$ and $q$ be connected through the
relationships
\begin{equation}
\label{qast} q^{\ast} = \frac{\alpha - 2(q-1)}{\alpha} \, \,
\mbox{and} \, \, q_{\ast} = \frac{\alpha q +  (q-1)}{\alpha + (q -
q).}
\end{equation}
Then the mapping
\begin{equation}
\label{mapping10} \mathcal{F}_q: \mathcal{G}_{q}[\alpha] \to
\mathcal{G}_{q_{\ast}}
\end{equation}
holds.
\end{proposition}

{\it Proof.} We use the scheme
\begin{align}
   &\mathcal{G}_q[\alpha]   \stackrel{\mathcal{F}_q}\longrightarrow
\mathcal{G}_{q_{\ast}}[\alpha] \notag \\
\label{scheme}
   \mathcal{M}_{q,q^{\ast}}  & \downarrow ~~\hspace{0.3in}  ~~ \uparrow \,\,
  \mathcal{K}^{-1}_{z(q^{\ast}),q_{\ast}} ~~
\\
 & \mathcal{G}_{q^{\ast}}[2]
\stackrel{F_{q^{\ast}}}\longrightarrow \mathcal{G}_{z(q^{\ast})}[2]
\notag
\end{align}
for the proof. Let a density $f \in \mathcal{G}_q [\alpha],$ i.e.
asymptotically $f(x) \sim C_f \, |x|^{- \alpha /(q-1)},
x\rightarrow\infty$ with some $C_f >0$. Its image
$\mathcal{M}_{q,q^{\ast}}[f](x),$  a $q^{\ast}$-Gaussian
$G_{q^{\ast}}(\beta; x),$ in order to be asymptotically equivalent
to $f,$ necessarily
\[
G_{q^{\ast}}(\beta; x) \sim \frac{C_1}{|x|^{2 \over {q^{\ast}-1}}}
\sim \frac{C_f}{|x|^{\alpha \over {q-1}}}, \, \, |x|\rightarrow
\infty.
\]
Hence,
\[
q^{\ast}= \frac{\alpha+2(q-1)}{\alpha} = 1 + \frac{2(q-1)}{\alpha}.
\]
Further, it follows from Corollary 2.10 of
\cite{UmarovTsallisSteinberg}, that
\[
F_{q^{\ast}}: \, \, \mathcal{G}_{q^{\ast}}[2] \rightarrow
\mathcal{G}_{q_1} [2],
\]
where
\[
q_1 = \frac{1+q^{\ast}}{3-q^{\ast}} = \frac{\alpha + (q-1)}{\alpha -
(q-1)}.
\]
Now taking into account the asymptotic equality (the right vertical
line in (\ref{scheme}))
\[
G_{q_1}(\beta_1; x) \sim \frac{C_2}{|x|^{2 \over {q_{1}-1}}} \sim
\frac{C_3}{|x|^{{\alpha} \over {q_{\ast}-1}} }, \, \, |x|
\rightarrow \infty,
\]
we obtain
\[
q_{\ast}=\frac{\alpha q - (q-1)}{\alpha - (q-1)} = 1 + \frac{\alpha
(q-1)}{\alpha  - (q-1)}.
\]
Thus, the mapping (\ref{mapping10}) holds with $q^{\ast}$ and
$q_{\ast}$ in equation (\ref{qast}). \eproof \vspace{.3cm}

Let us now introduce two functions that are important for our
further analysis:
\begin{align}
z_{\alpha}(s)&=\frac{\alpha s - (s-1)}{\alpha - (s-1)} = 1 +
\frac{\alpha(s-1)}{\alpha -(s-1)},
\end{align}
where $0<\alpha \le 2,$  $s < \alpha + 1,$ and
\begin{equation}
\zeta_{\alpha}(s) = \frac{\alpha + 2(s-1)}{\alpha} = 1 +
\frac{2(s-1)}{\alpha}, \, \,  0<\alpha \le 2 \,.
\end{equation}
It can be easily verified that $\zeta_{\alpha}(s)=s$ if $\alpha =
2.$

The inverse, $z_{\alpha}^{-1}(t), \, t \in (1 - \alpha, \infty)$, of
the the first function reads
\begin{equation}
z_{\alpha}^{-1}(t) =
         \frac{\alpha t + (t-1)}{\alpha + (t-1 )} = 1 +
\frac{\alpha(t-1)}{\alpha + (t-1)}\,.
\end{equation}
The function $z(s)$ possess the properties:
$z_{\alpha}(\frac{1}{z_{\alpha}(s)})= {1 \over s}$ and
$z_{\alpha}({1 \over s})={1 \over z_{\alpha}^{-1}(s)}\,.$ If we
denote $q_{\alpha, 1}=z_{\alpha}(q)$ and $q_{\alpha,
-1}=z_{\alpha}^{-1}(q),$ then
\begin{equation}
\label{qdual1} z_{\alpha}({1 \over q_{\alpha, 1}})={1 \over q} \, \,
\, \, \, \mbox{and} \, \, \, \, \, z_{\alpha}({1 \over q})={1 \over
q_{\alpha, -1}} \,.
\end{equation}

%
Proposition \ref{formula1} implies that for $0<\alpha \le 2$ and $1
\le q < \min \{2,1+\alpha\}$ the following mappings hold:
\begin{itemize}
\item[(i)]
$
\mathcal{F}_{q}:\mathcal{G}_q [\alpha] {\rightarrow}
\mathcal{G}_{z_{\alpha}(q)}[\alpha],
$
%
\item[(ii)]
$
\mathcal{F}^{-1}_{q}: \mathcal{G}_{z_{\alpha}(q)}[\alpha] \, \,
{\rightarrow} \, \, \mathcal{G}_q [\alpha],   \, \, \,
$
\end{itemize}
%
where $\mathcal{F}^{-1}_{q}$ is the inverse to $\mathcal{F}_{q}.$

It should be noted that as Hilhorst \cite{Hilhorst} noticed
$q$-Fourier transform in general is not one-to-one in the space of
densities. In paper \cite{UmarovTsallis} the invertibility of $F_q$
in the set of $q$-Gaussians is established. Since mappings
$\mathcal{M}_{q,q^{\ast}}$ and $\mathcal{K}_{q,q^{\ast}}$ are
one-to-one, relationship (\ref{scheme}) yields invertibility of
$\mathcal{F}_{q}$ in $\mathcal{G}_{z_{\alpha}(q)}[\alpha]$ and
validity of property $(ii).$

Further, we introduce the sequence $q_{\alpha,n} = z_{\alpha,n}(q) =
z(z_{\alpha,n-1}(q)), n=1,2,...,$  with a given $q = z_0(q), \,
q<1+\alpha.$ We can extend the sequence $q_{\alpha,n}$ for negative
integers $n=-1,-2,...$ as well, setting $q_{\alpha,-n} =
z_{\alpha,-n}(q)=z_{\alpha}^{-1}(z_{\alpha, 1-n}(q)), n = 1,
2,...\,.$ It is not hard to verify that
\begin{equation}
\label{qn} q_{\alpha, n} = 1 + \frac{\alpha (q-1)}{\alpha - n(q-1)}
= \frac{\alpha q - n(q-1)}{\alpha - n (q-1)},
\end{equation}
for all integer $n$ satisfying  $-\infty < n \le
[\frac{\alpha}{q-1}].$ The restriction $n\le [\alpha/(q-1)]$
implies the necessary condition $q_{\alpha,n}>1,$ since $q$-Fourier
transform is defined for $q \ge 1.$
Note that $q_{\alpha,n}$ is a function of $(q,n/\alpha)$, that
$q_{\alpha,n} \equiv 1$ for all $n=0, \pm 1, \pm 2,...,$ if $q=1$,
and that $\lim_{ n \rightarrow \pm \infty}z_{\alpha,n}(q)=1$ for all
$q \neq 1.$ Equation (\ref{qn}) can be rewritten as follows:
\begin{equation}
\frac{\alpha}{q_{\alpha,n}-1}-n=\frac{\alpha}{q-1} \,,      \,\,\,
n=0, \pm1, \pm 2,... \label{rewrite}
\end{equation}

We note that the latter coincides with equation (13) of
\cite{MendesTsallis2001}, once we identify  $\alpha$ with the
quantity $z$ therein defined, which was obtained through a quite
different approach (related to the renormalization of the index $q$
emerging from summing a specific expression over one degree of
freedom).
\par
We also note an interesting property of $q_{\alpha,n}$. If we have a
$q$-Gaussian in the variable $|x|^{\alpha/2}$ ($q \ge 1$), i.e., a
$q$-exponential in the variable $|x|^\alpha$, its successive
derivatives and integrations with respect to $|x|^\alpha$ precisely
correspond to $q_{\alpha,n}$-exponentials in the same variable
$|x|^\alpha.$

%
Further, we introduce the sequence
$q^{\ast}_{\alpha,n}=\zeta(q_{\alpha,n})$, which can be written in
the form
\begin{equation}
\label{qstarn} q^{\ast}_{\alpha,n}=1+\frac{2(q-1)}{\alpha - n(q-1)}
= \frac{\alpha + (n-2)(1-q)}{\alpha-n(q-1)},
\end{equation}
for $n =0,\pm 1,...,$ or, equivalently,
\begin{equation}
\label{qstarn2} \frac{2}{q^{\ast}_{\alpha,n}-1}+n=\frac{\alpha}{q-1}
 \, , \, \, n =0,\pm 1,....
\end{equation}
It follows from Proposition \ref{formula1} and definitions of
sequences $q_{\alpha,n}$ and $q^{\ast}_{\alpha, n}$ that
\begin{equation}
\label{formula2} \mathcal{F}_{q_{\alpha,n}}: \, \,
\mathcal{G}_{q_{\alpha,n}}[\alpha] \rightarrow
\mathcal{G}_{q_{\alpha,n+1}}, -\infty < n \le [\frac{\alpha}{q-1}].
\end{equation}
\begin{proposition}
\label{dual} For all $n=0, \pm1, \pm2,...$ the following relations
\begin{equation}
\label{dualityrelation} q^{\ast}_{\alpha, n-1} +
\frac{1}{q^{\ast}_{\alpha, n+1}} = 2,
\end{equation}
\begin{equation}
\label{p2n} q^{\ast}_{2,n}=q_{2,n} \, ,
\end{equation}
hold.
\end{proposition}
%

{\it Proof.} We notice that
\[
\frac{1}{q^{\ast}_{\alpha, n+1}} = 1 - \frac{2(q-1)}{\alpha -
(n-1)(q-1)}.
\]
On the other hand, by (\ref{qstarn})
\[
-\frac{2(q-1)}{\alpha - (n-1)(q-1)}=1-q_{\alpha,n-1}^{\ast},
\]
 which implies (\ref{dualityrelation})
immediately. The relation (\ref{p2n}) can be checked easily. \eproof

The property $q^{\ast}_{2,n}=q_{2,n}$ shows that the sequences
(\ref{qn}) and (\ref{qstarn}) coincide if $\alpha=2$. Hence, the
mapping (\ref{formula2}) takes the form $F_{q_{2,n}}: \, \,
\mathcal{G}_{q_{2,n}}(2) \rightarrow \mathcal{G}_{q_{2,n+1}}(2),$
recovering Lemma 2.16 of \cite{UmarovTsallisSteinberg}. Moreover, in
this case the duality (\ref{dualityrelation}) holds for the sequence
$q_{\alpha,n}$ as well. If $\alpha < 2$ then the values of
$q_{\alpha,n}^{\ast}$ are distinct  from the values of
$q_{\alpha,n}.$ The difference is given by
\[
q_{\alpha,n}-q^{\ast}_{\alpha,n}=\frac{(2-\alpha)(1-q)}{\alpha+n(1-q)},
\]
vanishing for $\alpha=2 \,, \forall q$, or for $q=1 \,, \forall
\alpha$. In the latter case $q_{\alpha,n}=q^{\ast}_{\alpha,n} \equiv
1.$

Further, we define for $n=0, \pm 1, ..., \, \,  k=1,2,...,$ $n+k \le
[\frac{\alpha}{q-1}]+1,$ the operators
\begin{align*}
\mathcal{F}^k_n (f) &=\mathcal{F}_{q_{\alpha, n+k-1}} \circ ...
\circ \mathcal{F}_{q_{\alpha, n}}[f]  \\
&=
\mathcal{F}_{q_{\alpha, n+k-1}}[...\mathcal{F}_{q_{\alpha,
n+1}}[\mathcal{F}_{q_{\alpha, n}}[f]]...],
\end{align*}
and
\begin{align*}
\mathcal{F}^{-k}_n (f)&=\mathcal{F}^{-1}_{q_{\alpha, n-k}} \circ ...
\circ \mathcal{F}^{-1}_{q_{\alpha, n-1}}[f] \\
&=
\mathcal{F}^{-1}_{q_{\alpha, n-k}}[...\mathcal{F}^{-1}_{q_{\alpha,
n-2}}[\mathcal{F}^{-1}_{q_{\alpha, n-1}}[f]]...].
\end{align*}
In addition, we assume that $\mathcal{F}_{q}^{k}[f]=f,$ if $k=0$ for
any appropriate $q.$
Summarizing the above mentioned relationships, we obtain the
following assertions.
\begin{proposition}
\label{lemma} The following mappings hold:
\begin{enumerate}
\item
$\mathcal{F}_{q_{\alpha,n}}: \, \,
\mathcal{G}_{q_{\alpha,n}}[\alpha] \, \, {\rightarrow} \,
\mathcal{G}_{q_{\alpha,n+1}}[\alpha], -\infty < n \le
[\frac{\alpha}{q-1}];$
\item
$\mathcal{F}^k_n:\mathcal{G}_{q_{\alpha,n}} [\alpha] \,
{\rightarrow} \, \mathcal{G}_{q_{\alpha,k+n}} [\alpha], \, \, \,
k=1,2,..., \, \, ~~ ~\\ n = 0, \pm1,..., -\infty < n+k \le
[\frac{\alpha}{q-1}]+1; $
\item
$\lim_{ k \rightarrow - \infty} \mathcal{F}^{k}_n \mathcal{G}_q
[\alpha] = \mathcal{G} [\alpha], \, \,  n = 0, \pm 1,...,$
\end{enumerate}
where $\mathcal{G}(\alpha)$ is the set of densities of classic
symmetric $\alpha$-stable L\'evy distributions.
\end{proposition}

\begin{theorem}  Assume $0<\alpha \leq 2$ and a sequence
$q_{\alpha,n}, \, -\infty < n \le [\alpha/(q-1)],$ is given as in
(\ref{qn}) with $q_0=q \in [1, \min \{2, 1+\alpha\}  ).$ Let $X_N$
be a symmetric $q_{\alpha, k}$-independent (for some $-\infty < k
\le [\alpha/(q-1)]$ and $\alpha \in (0,2])$) random variables all
having the same probability density function $f(x)$ satisfying the
conditions of Proposition \ref{mainlemma}.

Then the sequence
$$Z_N = \frac{X_1+...+X_N}
{(\mu_{q_{\alpha,k},\alpha}N)^{\frac{1}{\alpha (2-q_{\alpha,k})}}}
\, ,
$$ is $q_{\alpha,k}$-convergent
 to a
$(q_{\alpha,k-1},\alpha)$-stable distribution, as $N \rightarrow
\infty.$
\end{theorem}
\par
Proof. The case $\alpha=2$ coincides with Theorem 1 of
\cite{UmarovTsallisSteinberg}. For $k=0$, the first part of Theorem
($q$-convergence) is proved in Section 5 of the present paper. The
same method can be applied for $k \neq 1.$ For the readers
convenience we proceed the proof of the first part also in the
general case, namely for arbitrary $k.$ Suppose that $0<\alpha<2$.
We evaluate $F_{q_{\alpha,k}}(Z_N).$ Denote $Y_j =
X_j/{s_N(q_{\alpha,k})}, j=1,2,...,$ where ${s_N(q_{\alpha,k})}=
{(\mu_{q_{\alpha,k},\alpha}N)^{\frac{1}{\alpha (2-q_{\alpha,k})}}}.$
Then $Z_N = Y_1 +...+Y_N.$ Again using the relationship $F_q
[aX](\xi)=F_q[X](a^{2-q} \xi)$, we obtain
$F_{q_{\alpha,k}}(Y_1)=F_{q_{\alpha,k}}[f]( \frac{\xi}{
(\mu_{q_{\alpha,k},\alpha}N)^{\frac{1}{\alpha}} } ).$ Further, it
follows from  $q_{\alpha,k}$-independence of $X_1,X_2,...$ and the
associativity property of the $q$-product that
\begin{align}
\label{step100} &F_{q_{\alpha,k}}[Z_N](\xi) =
\otimes^N_{q_{\alpha,k}} F_{q_{\alpha,k}}[f]( \frac{\xi}{
(\mu_{q_{\alpha,k},\alpha}N)^{\frac{1}{\alpha}}  } ) , ~
\end{align}
the right hand side of which exhibits the $q_{\alpha,k}$-product of
$N$ identical factors $F_{q_{\alpha,k}}[f]( \frac{\xi}{
(\mu_{q_{\alpha,k},\alpha}N)^{\frac{1}{\alpha}}  } ).$ Hence, making
use of the properties of the $q$-logarithm, from (\ref{step100}) we
obtain
\begin{align}
\ln_{q_{\alpha,k}} F_{q_{\alpha,k}}[Z_N](\xi)&= N \ln_{q_{\alpha,k}}
F_{q_{\alpha,k}}[f]( \frac{\xi}{
(\mu_{q_{\alpha,k},\alpha}N)^{\frac{1}{\alpha}} } ) \notag \\
            &= N \ln_{q_{\alpha,k}} ( 1- \frac{|\xi|^{\alpha}}{N} +
o(\frac{|\xi|^{\alpha}}{N}))\notag \\
\label{step101}
& = - |\xi|^{\alpha} + o(1), \, N \rightarrow \infty
\,,
\end{align}
locally uniformly by $\xi$.
Consequently, locally uniformly by $\xi,$
\begin{equation}
\label{step_50} \lim_{N \rightarrow \infty} F_{q_{\alpha,k}}(Z_N) =
e_{q_{\alpha,k}}^{- |\xi|^{\alpha}} \in \mathcal{G}_{q_{\alpha,k}}
(\alpha) \,.
\end{equation}
Thus, $Z_N$ is $q_{\alpha,k}$-convergent.
\par
To show the second part of Theorem we use Proposition \ref{lemma}.
In accordance with this lemma there exists a density $f(x) \in
\mathcal{G}_{q_{\alpha,k-1}}[\alpha],$ such that
$\mathcal{F}_{_{\alpha,k-1}}[f] = e_{q_{\alpha,k}}^{-
|\xi|^{\alpha}}.$ Hence, $Z_N$ is $q_{\alpha,k}$-convergent to a
$(q_{\alpha,k-1},\alpha)$-stable distribution, as $N \rightarrow
\infty.$ \eproof

\section{Scaling rate analysis}
In paper \cite{UmarovTsallisSteinberg} the formula
\begin{equation}
\label{betak} \beta_k = \Bigl(\frac{3-q_{k-1}}{4 q_k C_{q_{k-1}}^{2
q_{k-1} -2}}\Bigr)^{1 \over {2-q_{k-1}}}.
\end{equation}
was obtained for the $q$-Gaussian parameter $\beta$ of the
attractor. It follows from this formula that the scaling rate in the
case $\alpha = 2$ is
\begin{equation}
\delta=\frac{1}{2-q_{k-1}} = q_{k+1},
\end{equation}
where $q_{k-1}$ is the $q$-index of the attractor. Moreover, if we
insert the 'evolution parameter' $t$, then the translation of a
$q$-Gaussian to a density in $\mathcal{G}_q[\alpha]$ changes $t$ to
$t^{2/\alpha}.$ Hence, applying these two facts to the general case,
$0<\alpha \le 2,$ and taking into account that the attractor index
in our case is $q^{\ast}_{\alpha, k-1}$, we obtain the formula for
the scaling rate
\begin{equation}
\delta=\frac{2}{\alpha (2-q^{\ast}_{\alpha, k-1})}.
\end{equation}
In accordance with Proposition \ref{dual},  $2-q^{\ast}_{\alpha,
k-1} = 1/q^{\ast}_{\alpha, k+1}.$ Consequently,
\begin{equation}
\label{scaling1} \delta=\frac{2}{\alpha} q^{\ast}_{\alpha, k+1} =
\frac{2}{\alpha} \frac{\alpha - (k-1)(q-1)}{\alpha - (k+1)(q-1)}.
\end{equation}
Finally, in terms of $Q=2q-1$ the formula (\ref{scaling1}) takes the
form
\begin{equation}
\label{scaling2} \delta=\frac{2}{\alpha} \frac{2 \alpha -
(k-1)(Q-1)}{2 \alpha - (k+1)(Q-1)}.
\end{equation}

In \cite{UmarovTsallisSteinberg} it was noticed that the scaling
rate in the non-linear Fokker-Planck equation can be derived from
the model corresponding to the case $k=1.$ Taking this fact into
account we can {\it conjecture that the scaling rate in the
fractional generalization of the nonlinear Fokker-Planck equation is
$$\delta=\frac{2}{\alpha +1 - Q},$$} which can be derived from
(\ref{scaling2}) setting $k=1$. In the case $\alpha=2$ we get the
known result $\delta = 2/(3-Q)$ obtained in \cite{TsallisBukman}.

\section{On additive and multiplicative dualities}

In the nonextensive statistical mechanical literature, there are two
transformations that appear quite frequently in various contexts.
They are sometimes referred to as {\it dualities}.  The {\it
multiplicative duality} is defined through
\begin{equation}
\mu (q)=1/q \,,
\end{equation}
and the {\it additive duality} is defined through
\begin{equation}
\nu (q)=2-q \,.
\end{equation}
They satisfy $\mu^2=\nu^2={\bf 1}$, where $\bf 1$ represents the
{\it identity}, i.e., ${\bf 1}(q)=q, \forall q$. We also verify that
\begin{equation*}
(\mu\nu)^m(\nu\mu)^{m}= (\nu\mu)^m(\mu\nu)^{m}={\bf
1}\;\;\;\;(m=0,1,2,...) \,.
\end{equation*}
Consistently, we define $(\mu\nu)^{-m} \equiv (\nu\mu)^{m}$, and
$(\nu\mu)^{-m} \equiv (\mu\nu)^{m}$ .

Also, for $m=0,\pm 1, \pm2, ...$, and $\forall q$,
\begin{equation}
\label{duality} (\mu\nu)^m(q) = \frac{m-(m-1)\,q}{m+1-m\,q}
=\frac{q+m(1-q)}{1+m(1-q)} \,,
\end{equation}
\begin{equation*}
\nu(\mu\nu)^m(q) = \frac{m+2-(m+1)\,q}{m+1-m\,q}
=\frac{2-q+m(1-q)}{1+m(1-q)} \,,
\end{equation*}
and
\begin{equation*}
(\mu\nu)^m\mu(q) = \frac{-m+1+m\,q}{-m+(m+1)\,q}
=\frac{1-m(1-q)}{q-m(1-q)} \,.
\end{equation*}

We can easily verify, from equations (\ref{qn}) and (\ref{duality}),
that the sequences $q_{2,n}$ ($n=0,\pm 2,\pm 4, ...$) and $q_{1,n}$
($n=0,\pm 1,\pm 2, ...$) coincide with the sequence $(\mu\nu)^m(q)$
($m=0, \pm 1,\pm ,2,...$).

\section{Classification of $(q,\alpha)$-stable distributions and some conjectures}

The $q$-CLT formulated in \cite{UmarovTsallisSteinberg} states that
the appropriately scaling limit of sums of $q_k$-independent random
variables with a finite $(2q_k-1)$-variance is a
$q^{\ast}_k$-Gaussian, which is a $q^{\ast}_k$-Fourier preimage of a
$q_k$-Gaussian. Here $q_k$ and $q^{\ast}_k$ are sequences defined as
\[
q_k=\frac{2q-k(q-1)}{2-k(q-1)}, \, k = 0, \pm 1, ...,
\]
and
\[
q_k^{\ast}=q_{k-1}, \, k=0, \pm 1, ....
\]
Schematically $q$-CLT in \cite{UmarovTsallisSteinberg} can be
represented as
\begin{equation}
\label{schemeCLT} \{f:\sigma_{2q_k-1}(f)<\infty\}
\stackrel{F_{q_k}}{\longrightarrow} \mathcal{G}_{q_k}[2]
\stackrel{F_{q^{\ast}_{k}}}{\longleftarrow}
\mathcal{G}_{q_k^{\ast}}[2].
\end{equation}
We have also noticed that $q$-CLT can be  described by the triplet
$(P_{att},P_{cor},P_{scl})$, where $P_{att}, \, P_{cor}$ and $
P_{scl}$ represent parameters of {\it the attractor}, {\it the
correlation} and {\it the scaling rate}, respectively. We found that
(see details in \cite{UmarovTsallisSteinberg}) for $q$-CLT this
triplet
\begin{equation}
\label{triplet} (P_{att},P_{cor},P_{scl}) \equiv (q_{k-1}, q_k,
q_{k+1}).
\end{equation}
\par

Schematically Theorem 3 of the current paper can be represented as
\begin{equation}
\label{schemeLevy1}
 \mathcal{L}_q[\alpha]
\stackrel{F_{q}}{\longrightarrow} \mathcal{G}_{q}[\alpha]
\stackrel{F_q}{\longleftarrow}  \mathcal{G}_{q^{L}} [2], \,
0<\alpha<2,
\end{equation}
where $\mathcal{L}_q[\alpha]$ is the set of $(q,\alpha)$-stable
distributions, $\mathcal{G}_{q^{L}} [2]$ is the set of
$q^L$-Gaussians with index $q^L$ defined as
\[
q^L = q^{L}_{\alpha}(q) = \frac{3+(2q-1)\alpha}{1+\alpha}.
\]
Recall that the case $\alpha=2$ was peculiar and we agree to refer
to the scheme (\ref{schemeCLT}) in this case.
\par
Theorem 4 generalizes the $q$-CLT (which corresponds to $\alpha=2$)
to the whole range $0<\alpha \le 2$. Schematically this theorem can
be represented as
\begin{equation}
\label{schemeLevy2} \mathcal{L}_{q_{\alpha,k}}[\alpha]
\stackrel{F_{q_{\alpha,k}}}{\longrightarrow}
\mathcal{G}_{q_{\alpha,k}}[\alpha]
\stackrel{F_{q^{\ast}_{\alpha,k}}}{\longleftarrow}
\mathcal{G}_{q_{\alpha,k}^{\ast}}[2], \, 0<\alpha \le 2,
\end{equation}
generalizing the scheme (\ref{schemeCLT}). The sequences
$q_{\alpha,k}$ and $q^{\ast}_{\alpha,k}$ in this case read
\[
q_{\alpha,k}=\frac{\alpha q+k(1-q)}{\alpha+k(1-q)}, \, k = 0, \pm 1,
...,
\]
and
\[
q_{\alpha,k}^{\ast}= 1 - \frac{2(1-q)}{\alpha + k (1-q)}, \, k=0,
\pm 1, ....
\]
Note that the triplet $(P_{att},P_{cor},P_{scl})$ mentioned above,
in this case, takes the form
\begin{equation} \label{triplet1}
(P_{att},P_{cor},P_{scl}) \equiv (q^{\ast}_{\alpha,k-1}, \,
q_{\alpha,k},\, (2/\alpha) q^{\ast}_{\alpha,k+1}),
\end{equation}
recovering the triplet (\ref{triplet}) in the case $\alpha = 2.$
\par
In connection with the above discussion about triplets, we note that
the existence of a $q$-triplet, namely $(q_{sen},q_{rel},q_{stat})$,
related respectively to sensitivity to the initial conditions,
relaxation, and stationary state was conjectured in
\cite{Tsallistriplet}. Later it was observed in the solar wind at
the distant heliosphere \cite{BurlagaVinas,BurlagaNessAcuna}. The
triplet in (\ref{triplet}) obtained theoretically might be useful
hint for its understanding.

Finally, unifying the schemes (\ref{schemeLevy1}) and
(\ref{schemeLevy2}) we obtain the general picture for the
description of $(q,\alpha)$-stable distributions:
\begin{equation}
\label{schemeLevy3} \mathcal{L}_{q_{\alpha,k}}[\alpha]
\stackrel{F_{q_{\alpha,k}}}{\longrightarrow}
\mathcal{G}_{q_{\alpha,k}}[\alpha]
\stackrel{F_{q^{\ast}_{\alpha,k}}}{\longleftrightarrow}
\mathcal{G}_{q_{\alpha,k}^{\ast}}[2]
\end{equation}
\hspace{1.5in} $\updownarrow \, F_{q}$
\[
\hspace{0.1in} \mathcal{G}_{q^{L}_{\alpha,k}} [2],
\]
where
\[
 q^{L}_{\alpha,k} = q^{L}_{\alpha}(q_{\alpha,k}) = \frac{3+(2q_{\alpha, k}-1)\alpha}{1+\alpha} \,.
\]

\begin{figure}[hpt]
\vspace{-0.3cm} \centering
\includegraphics[width=4.1in]{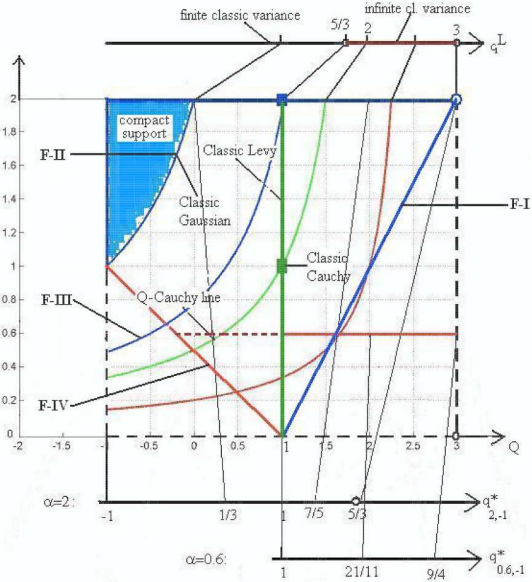}
\vspace{-3.5cm} \caption{\scriptsize $(Q,\alpha)$-regions. }
\label{fig:conclusion}
\end{figure}

In Fig. \ref{fig:conclusion} the dependence of  $q^L$ and $q^{\ast}$
on parameters $(Q,\alpha) \in \mathcal{Q}$ in the case $k=0$ is
represented. If $Q = 1$ and $\alpha = 2$ (the blue box in the
figure), then the random variables are independent in the usual
sense and have {\it finite} variance. The standard CLT applies, and
the attractors are classic Gaussians.

If $Q$ belongs to the interval $(1, 3)$ and $\alpha = 2$ (the blue
straight line on the top), the random variables are {\it not}
independent. If the random variables have a {\it finite} Q-variance,
then $q$-CLT \cite{UmarovTsallisSteinberg} applies, and the
attractors belong to the family of $q^{\ast}$-Gaussians. Note that
$q^{\ast}$ runs in $[1, 5/3).$ Thus, in this case, attractors
($q^{\ast}$-Gaussians) have {\it finite} classic variance (i.e.,
$1$-variance) in addition to {\it finite} $q^{\ast}$-variance.

If Q = 1 and $0 < \alpha <2$ (the vertical green line in the
figure), we have the classic L\'evy distributions, and random
variables are independent, and have {\it infinite} variance. Their
scaling limits-attractors belong to the family of $\alpha$-stable
L\'evy distributions. It follows from (\ref{Q=1}) that in terms of
$q$-Gaussians classic symmetric $\alpha$-stable distributions
correspond to $\cup_{5/3<q<3} \mathcal{G}_q$
\cite{UmarovTsallis2007}.

If $0 < \alpha < 2,$ and $Q$ belong to the interval $(1, 3)$ we
observe the rich variety of possibilities of $(q,\alpha)$-stable
distributions. In this case random variables are {\it not}
independent, have {\it infinite} variance and {\it infinite
Q-variance}. The rectangle $\{1<Q<3; \, \, 0<\alpha<2 \}$, at the
right of the classic L\'evy line, is covered by non-intersecting
curves
$$
C_{q^L} \equiv \{(Q,\alpha): \frac{3+Q\alpha}{\alpha + 1} = q^L\},
\, 5/3<q^L<3 \,.
$$
This family of curves describes all $(Q,\alpha)$-stable
distributions based on the mapping (\ref{schemeLevy1}) with
$q$-Fourier transform. The constant $q^L$ is the index of the
$q^L$-Gaussian attractor corresponding to the points $(Q,\alpha)$ on
the curve $C_{q^L}$. For example, the green curve corresponding to
$q^L=2$ describes all $Q$-Cauchy distributions, recovering the
classic Cauchy-Poisson distribution if $\alpha = 1$ (the green box
in the figure). Every point $(Q,\alpha)$ lying on the brown curve
corresponds to $q^L=2.5$.

The second classification of $(Q,\alpha)$-stable distributions
presented in the current paper, and based on the mapping
(\ref{schemeLevy2}) with $q^{\ast}$-Fourier transform leads to a
covering of $\mathcal{Q}$ by curves distinct from $C_{q^L}$. Namely,
in this case we have the following family of straight lines
\begin{equation}
\label{lines} L_{q^{\ast}} \equiv \{(Q,\alpha): \frac{4 \alpha}{Q +
2 \alpha - 1} = 3 - q^{\ast} \}, \, 1 \le q^{\ast} <3,
\end{equation}
which are obtained from (\ref{qstarn}) replacing $n=-1$ and
$2q-1=Q.$ For instance, every $(Q,\alpha)$ on the line F-I (the blue
diagonal of the rectangle in the figure) identifies
$q^{\ast}$-Gaussians with $q^{\ast}=5/3$. This line is the frontier
of points $(Q,\alpha)$ with finite and infinite classic variances.
Namely, all $(Q,\alpha)$ above the line F-I identify attractors with
{\it finite} variance, and points on this line and below identify
attractors with {\it infinite} classic variance. Two bottom lines in
Fig. \ref{fig:conclusion} reflect the sets of $q^{\ast}$
corresponding to lines $\{1\le Q<3; \alpha = 2)\}$ (the top boundary
of the rectangle in the figure) and $\{1\le Q<3; \alpha = 0.6)\}$
(the brown horizontal line in the figure).

\vspace{.2cm}

{\bf Some conjectures.} Both classifications of $(Q,\alpha)$-stable
distributions are restricted to the region $Q=\{1\le Q<3, 0<\alpha
\le 2\}.$ This limitation is caused by the tool used for these
representations, namely, $Q$-Fourier transform is defined for $Q\ge
1.$ However, at least two facts, the positivity of $\mu_{q,\alpha}$
in Proposition \ref{mainlemma} for $q>\max \{0,1-1/\alpha\}$ (or,
the same, $Q>\max \{-1, 1-2/\alpha\}$) and continuous extensions of
curves in the family $C_{q^L}$, strongly indicate to following
conjectures, regarding the region $\{Q<1\}$ on the left to the
vertical green line (the classic L\'evy line) in Fig.
\ref{fig:conclusion}. In this region we see three frontier lines,
F-II, F-III and F-IV.

\begin{conjecture} The line F-II splits the regions where the
random variables have {\it finite} and {\it infinite} $Q$-variances.
More precisely, the random variables corresponding to $(Q,\alpha)$
on and above the line F-II have a {\it finite} $Q$-variance, and,
consequently, $q$-CLT \cite{UmarovTsallisSteinberg} applies.
Moreover, as seen in the figure, the $q^L$-attractors corresponding
to the points on the line F-II are the classic Gaussians, because
$q^L=1$ for these $(Q,\alpha)$. It follows from this fact, that
$q^L$-Gaussians corresponding to points above F-II have compact
support (the blue region in the figure), and $q^L$-Gaussians
corresponding to points on this line and below have infinite
support.
\end{conjecture}

\begin{conjecture}
The line F-III splits the points $(Q,\alpha)$ whose $q^L$-attractors
have {\it finite} or {\it infinite} classic variances. More
precisely, the points $(Q,\alpha)$ above this line identify
attractors (in terms of $q^L$-Gaussians) with {\it finite} classic
variance, and the points on this line and below identify attractors
with {\it infinite} classic variance.
\end{conjecture}

\begin{conjecture}
The frontier line F-IV with the equation $Q + 2 \alpha -1 = 0$ and
joining the points $(1,0)$ and $(-1,1)$ is related to attractors in
terms of $q^{\ast}$-Gaussians. It follows from (\ref{lines}) that
for $(Q,\alpha)$ lying on the line F-IV, the index
$q^{\ast}=-\infty$. Thus the horizontal lines corresponding to
$\alpha < 1$ can be continued only up to the line F-IV with
$q^{\ast} \in (-\infty, 3-\frac{4 \alpha}{Q + 2 \alpha - 1})$ (see
the dashed horizontal brown line in the figure). If $\alpha
\rightarrow 0,$ the $Q$-interval becomes narrower, but
$q^{\ast}$-interval becomes larger tending to $(-\infty, 3)$.
\end{conjecture}

Results confirming or refuting any of these conjectures would be an
essential contribution to deeper understanding of the nature of
$(Q,\alpha)$-stable distributions, and nonextensive statistical
mechanics, in particular.

\vspace{.2cm}

Finally we note that  Fig. \ref{fig:conclusion} corresponds to the
case $k=0$ in the description (\ref{schemeLevy3}). The cases $k \neq
0$ can be treated in the same way. \\

{\bf Acknowledgement.} We acknowledge thoughtful remarks by R.
Hersh, E.P. Borges and S.M.D. Queiros. Financial support by the
Fullbright Foundation, SI International and NIH grant P20
GM067594 (USA Agencies), and CNPq and Faperj (Brazilian Agencies) are
acknowledged as well.


%
%

\end{document}